# Scholarly Communication and
# the Continuum of Electronic Publishing

**Rob Kling and Geoffrey McKim**

v5.7
March 22, 1999

Center for Social Informatics
Indiana University School of Library and Information Science
Bloomington, IN 47405
www.slis.indiana.edu/CSI

Contact info: *kling@indiana.edu, mckimg@indiana.edu*

# ABSTRACT

Electronic publishing opportunities, manifested today in a variety of electronic journals and Web-based compendia, have captured the imagination of many scholars.  These opportunities have also destabilized norms about the character of legitimate scholarly publishing in some fields. Unfortunately, much of the literature about scholarly e-publishing homogenizes the character of publishing. This article provides an analytical approach for evaluating  disciplinary conventions and for proposing policies about scholarly e-publishing. We characterize three dimensions of scholarly publishing as a communicative practice -- publicity, access, and trustworthiness, and examine several forms of paper and electronic publications in this framework. This analysis shows how the common claim that e-publishing "substantially expands access" is over-simplified.  It also indicates  how peer-reviewing (whether in paper or electronically) provides valuable functions for scholarly communication that are not effectively replaced  by self-posting articles in electronic media.

Acknowledgements: This article was significantly improved by comments on intermediate drafts by Charles Bailey, Howard Becker, Holly Crawford, Blaise Cronin, Steve Harter, Don Kraft, Robin Peek, Tefko Seracevic, and John Walsh. Funding was provided in part by NSF grant #SBR-9872961.

# INTRODUCTION

Scholarly publishing practices – especially those related to electronic publishing – are rapidly evolving and have not yet formed stable configurations. The cacophonous discourse about electronic publishing as a means of scholarly communication can be thematized around a series of issues that "are in the talk" as much as "in the published literature": legitimacy of electronic publishing and electronic journals; whether electronic publishing will reduce overall costs, expand access to documentary materials, and in that sense democratize; whether scholars should vigorously embrace electronic publishing; how electronic publishing will affect traditional paper publishers; and how electronic publishing will affect research libraries.

Disciplinarity socializes junior scholars into beliefs about appropriate forms of scholarly communication, such as the relative value of working papers, conference papers, journals, anthologies and monographs. Editorial elites in the disciplines articulate conventions for participating scholars, especially those who are relatively junior. Today, there is significant diversity across some physics fields in which working papers are a legitimate communication format, computer science (which supports electronic publications that precede paper publication), and both American psychology and chemistry where the American Psychological Association (APA) and the American Chemical Society (ACS) have at times tried to ban scholarly electronic publishing (except in the form of society-sponsored electronic versions of existing paper publications). Scholars who work at the intersection of disciplines with differing publication practices are in a special bind.

Unfortunately, few analyses of scholarly e-publishing explicitly acknowledge the differences in communication practices from field to field. Terms like "being published" are treated as categorical. However, the actual communicative practices that constitute publishing vary from one field to another. For example, computer scientists often treat conference articles as significant forms of publication, and computer science journals are more likely to republish amplified versions of a conference article. In contrast, natural scientists insist that journal articles are the primary form of significant publication, and their best journals do not publish amplified versions of articles that have previously been published in very obscure journals.

This article provides an analytical approach for comparing disciplinary conventions. It also serves as basis for proposing policies about e-publishing, based on whether and how they improve in scholarly communication. It represents our sustained reflection on a variety of data, including the debates surrounding the future of journals in an increasingly electronically mediated environment, and the electronic publishing practices and policies of journals and scholarly societies, as derived from documentary evidence. We also use some data from interviews with active scholars in six disciplines about their electronic media use we conducted as part of a different study. However, we are not basing our analysis upon one systematically analyzed data set. Our primary contribution is to help clarify conceptually the relationships between various forms of paper and electronic publication. It helps to open new questions, as well as some new lines for systematic empirical inquiry.

## Definitions

It is useful to begin by clarifying several concepts related to electronic publishing. We define an **electronic publication** as a document distributed primarily through electronic media. The distribution medium is the defining factor, since an electronic publication may well be printed to be read, and may be circulated post-publication in printed form. Conversely, most scholarly publications distributed in paper form were produced on personal computers, and even typeset using software. According this definition, an article posted on a Web page (under a variety of restrictions or conditions), an article distributed via e-mail, or via an e-mail-based distribution list are all electronic publications.

*Varieties of Electronic Journals*

In discussions of the scholarly communication system, the term "electronic journal" is often characterized in conflicting ways. Consider the following four different glosses on the term. Machovec (1997) provides as examples of electronic journal projects Project Muse at Johns Hopkins University Press, the Journal Storage, Project (JSTOR), Elsevier Press, Springer-Verlag, Blackwell, Science, Highwire Press at Stanford University and Academic Press. All of his examples of electronic journals represent publishers or aggregators who are delivering conventional paper scholarly journals and an electronic version in parallel. On the other hand, MIT Press's Janet Fisher (1996) writes that "In the period from 1993 to 1995, the number of e-journals has increased, but they are still almost entirely free and created almost entirely by dedicated groups of individuals without production subsidy from institutions or scholarly associations (p. 231) ," referring primarily to journals or journal-like publications that are distributed in electronic form only, like *Bryn Mawr Classical Review, Postmodern Culture,* and *Psycholoquy* . Odlyzko (1996) proposes still a different view of electronic journals, as collections of unpackaged, but potentially refereed documents, available for download from a central server, akin to Paul Ginsparg's working article server:

> "The new technologies, however, are making possible easy publication of electronic journals by scholars alone. It is just as easy for editors to place manuscripts of refereed papers in a publicly accessible directory or preprint server as it is for them to do the same with their own preprints. The number of electronic journals is small, but it is rising rapidly (Odlyzko, 1996, p. 95)."

An environmental biologist colleague (Kling & McKim, 1997) held yet a different view. When asked if biologists would ever accept electronic journals, he replied with a categorical "no, not at all." When he was prompted with the question, "Even if they were peer-reviewed?", he revised his answer: "Oh, yes, of course they will if they are peer-reviewed." He had originally assumed that "electronic journals" were not peer-reviewed; he saw them as collections of self-published articles.

One consequence of this lack of consensus on the meaning of the term "electronic journal" is widely varying estimates of the number of active electronic journals. A second, more serious consequence is the potential for misunderstanding and miscommunicating changes in the use of electronic media for scholarly communication. For example, a report that scientists are increasingly searching the electronic archives provided by conventional journal publishers in order to locate articles, could be misinterpreted as evidence that scientists are increasingly using electronic journals in their research. However, such a report would not mean that scientists are using kind of electronic journals characterized by Odlyzko, or Fisher (Harter, 1998), since many publishers are providing electronic archives of journals that are primarily distributed in paper.

Some of these confusions can be avoided by using more precise definitions. We define an **e-journal** as an edited package of articles that is *distributed* to most of its subscribers in electronic form (Kling and Covi, 1995). Articles from an e-journal may, and probably will, be printed for careful reading, and might be stored in libraries in a printed form, for archival purposes. However, e-journals are accessed *primarily* in electronic form[1]. Examples of journals fitting into our definition include *Psycholoquy, MISQ Discovery, the Journal of the Association for Information Systems (JAIS), the World Wide Web Journal of Biology*, and the *Internet Journal of Science: Biological Chemistry*. Today there are remarkably few e-journals in the sciences, and they publish so few articles each year as to have minimal impact on scientific communication systems. For example, the *Internet Journal of Science: Biological Chemistry* published 8 articles in 1997 (along with some conference information). In the same year, 1997, the *Journal of Biological Chemistry* had published approximately 3500 articles.

While e-journals correspond to conventional paper journals, other forms of electronic publishing correspond with various paper-based forums. E-journals may or may not be peer-reviewed: "magazine" or "bulletin" are terms that characterize some forms of edited but non-refereed publications. Electronic scholarly communications that are not peer-reviewed are given a variety of labels, including e-prints, working papers, electronic magazines, and electronic newsletters. We refer to articles made publicly and electronically available in non-peer-reviewed form, either as posted on an individual or organizational Web page, or on a server such as the Los Alamos Physics E-Print Archive (http://www.arxiv.org/), as **electronic working articles**.

We contrast the e-journal with the **hybrid paper-electronic journal** (or **p-e journal**). The p-e journal is a package of peer-reviewed articles available through electronic channels, but whose primary distribution channels are paper-based. Examples of p-e journals include: *Science Online*, *Cell*, *Nature*, the *Journal of Biological Chemistry*, *Astrophysical Journal*, and the *Journal of Neuroscience.* The criteria for distinguishing an e-journal from a p-e journal are anchored in the readership. A p-e journal could certainly become an e-journal, if its readership changed their behaviors and made electronic access to the journal their primary means, and if paper copies were printed primarily for archives or libraries.

---

[1] Harter and Kim (1996) referred to this type of publication as a "pure electronic journal".

We also identify another type of hybrid – an **e-p journal**. An e-p journal -- a hybrid electronic-paper journal – is primarily distributed electronically, but may have limited distribution in paper form. The *Electronic Transactions on Artificial Intelligence* (*ETAI*) was designed to function primarily as an electronic journal. For example, *ETAI's* web site has a public discussion section linked to each submitted article. However, an annual paper edition of the articles, without the discussion, is also published by the Royal Swedish Academy of Sciences (KVA). This dual publication makes the *ETAI* into an e-p journal[2].

Unfortunately, few analysts distinguish between these three forms of electronic publications. Our p-e journals are often called simply "e-journals." Many scientific paper journals are becoming p-e journals (Association of Research Libraries, 1997). We suspect that reports of exponential growth of e-journals really mean exponential growth of p-e journals. This is not a minor matter, since the p-e journals bring their reputations, review practices that they established in the paper world, and some of their readership to their electronic versions. In contrast, new e-journals and e-p journals face more daunting problems in establishing their legitimacy, and risk a higher failure rate.

### Tensions in Scholarly Publishing

Our policy analysis was motivated by three tensions in the discourse surrounding electronic scholarly publishing. The first is a tension between a set of claims made by an emerging electronic publishing professional reform movement – claims that tend to maximize the advantages of electronic publishing and minimize the costs – and the realities experienced by many users of electronic documents and technologies.

The second lies in the variety one can find in the Internet publishing policies and practices of different scholarly societies and journals. Some of them strongly homogenize all forms of Internet posting, declaring them essentially equivalent to publishing (for the purposes of subsequent journal publication), while others provide authors with more leeway with which to maintain parallel copies of documents on personal or organizational Web sites. We will discuss four different sets of Internet posting practices from four different fields.

The third is the tension between the common assumption that posting an article on the Internet automatically and naturally ensures an appropriate scholarly audience or audiences for it, and the reality, in which scholarly communities are frequently bound together by, and structured by institutions (disciplines and disciplinary societies) and institutional circuits (such as journal subscriptions, conference mailing lists, conferences, journal referee lists, and "invisible colleges"). This tension can be translated into the question that drives our interest in electronic media and scholarly communication: *how can scholarly communication be improved using electronic media*

---

[2] See Kling (1999) for an extended discussion of *ETAI*'s design and functioning as an electronic journal.

*without undermining the useful functions currently provided by academic journals, including not only peer review, but editorial screening, manuscript solicitation, distribution, etc.*

Some electronic publishing enthusiasts, such as Odlyzko (1996), consider traditional printed journals "awkward artifacts" that will "likely disappear within 10 to 20 years." In contrast, we view both electronic and paper journals as fulfilling a set of useful communicative functions. Each medium provides a package that is convenient for some purposes, and awkward for others (i.e., low cost searching and distribution for e-journals, and ease of reading for paper journals). The extent to which e-journals or e-p journals will replace p-journals or p-e journals keenly concerns many scholarly societies and journal editors.

*The Electronic Publishing Reform Movement*

Electronic publishing is not simply a set of professional practices; it is also the focus of a small e-publishing professional reform movement. This reform movement shares much in common with other computerization movements (Kling & Iacono, 1995; Iacono & Kling, 1996). Like any professional reform movement, it is energized by a core group of energetic articulate activists, and is organized around some common reforms and an ideology. Also, many professionals and scholars can agree with some of the reforms (advocate some form of e-publishing) without being an active member of the movement.

This movement's **core group of enthusiasts** (e.g. Paul Ginsparg, Stevan Harnad, Andrew Odlyzko, and Ann Okerson) are well known for their provocative writings about e-publishing. Harnad, for example, is also known as the editor of the electronic journal *Psycholoquy*, as the originator of "scholarly skywriting," a short, discursive, and iterative form of scholarly communication (Harnad, 1991), and for his "subversive proposal," a radically decentralized scholarly publishing model, in which scholars self-publish their works, which then may or may not be peer-reviewed (Harnad, 1995; Brent, 1995). Ginsparg is best known as the developer of the Los Alamos National Labs Physics E-Print Archive, a working article server used by high-energy physicists (http://www.arxiv.org/).

Second, the movement's arguments are anchored in the precept that "electronic media are almost always better than paper." This position is arguable, but is often treated as a dogma, and based on several claims: electronic publishing is dramatically less expensive than paper publishing; access to electronic publications is easier and wider; and electronic publishing can speed up scientific communication. It is interesting to note that both Harnad's "subversive proposal" and Ginsparg's E-Print server bypass peer-review (although Harnad also values peer-review and discusses a way of augmenting his "subversive proposal" to include peer-review) (Brent, 1995).

The electronic publishing reform movement has been an interesting source of tension by articulately raising fundamental questions about the costs and efficacy of the paper-based system of scholarly communication. Like all professional reform movements, its participants raise issues that raise important issues that many professionals would prefer to ignore. But its stance is more problematic when its enthusiasts claim that a single model for electronic scholarly publishing is appropriate for all scholarly communities ("One size fits all"). These electronic publishing enthusiasts not only advocate the virtues of electronic publishing systems, but they also dominate the more visible discourses, and thus set expectations for the potentials of electronic publishing in varied scholarly communities, not just among those who are keenly interested in e-publishing reforms (see Kling and Lamb, 1996)[3].

*Internet Posting as Prior Publication: Practices and Policies*

There is a wide discrepancy between various professional societies in their stances towards the posting of documents on the Internet. These stances are reflected most clearly in their formal policies about prior publication. Prior publication policies indicate to what degree is some form of distribution, such as publishing in a paper conference proceedings, treated as "prior publication" by the editor of society journals. Today, one new and fundamental issue that such policies address is the status of articles that have been "posted on the Internet."

We will examine the wide disciplinary variation in prepublication policies, by examining recent policies in four fields: psychology, chemistry, computer science and information systems. These policies illustrate a range of approaches and indicate different ways of addressing the character of prior publication. The policies of many journals or scholarly societies are in flux, and there are also differences within specific fields.

One of the most widely publicized Internet publication policies comes from the American Psychological Association (1996). The APA's 1996 "interim policy," (revised in 1997), notes:[4]

> "Authors are instructed not to put their manuscripts on the Internet at any stage (draft, submitted for publication, in press, or published). Authors should be aware that they run a risk of having (a) their papers stolen, altered, or distributed without their permission and, very importantly, (b) an editor regard such papers as previously "published" and not eligible as a submission-a position taken by most APA journal editors.
>
> In addition, after acceptance for publication, the publisher is the copyright holder. APA does not permit authors to post the full text of their APA-published papers on the Internet at this time, as developments in the online world cannot be predicted. The APA will, however, closely follow such Internet developments. The P&C Board will establish a task force in June 1997 to investigate developments and recommend a longer term APA policy."

---

[3] Hafner (1998), for example, published an article in the New York Times that made it appear as if Paul Ginsparg singlehandedly conceived of the physics e-print server and a service that is easy to design, set up and operate effectively. Hafner quotes Harnad: "Sooner or later someone is going to be shrewd and prophetic enough to realize that Paul has quietly done something absolutely monumental .... When the historians write the history of it all, Paul Ginsparg will certainly get the full credit for having shown the way, not just by pointing, but by actually constructing the ultimate solution." We view Ginsparg's innovation as much more incremental, since it built on practices of widely sharing paper working papers among high energy physicists that was developed in the 1960s, as well as sharing abstracts working-paper abstracts on-line in the 1970s.

[4] In 1997 the APA(1997) moderated its stance on Web posting prior to publication in a society journal. The November 1997 policy gave journal editors explicit discretion over whether or not to accept articles that have been posted on the Internet for publication: "Such posted or shared documents may or may not be considered 'publications' by a given journal or editor, depending on the circumstance of the posting and the nature or orientation of the journals." The revised APA policy was not as chilling as the 1996 policy, but instead abdicated any responsibility to forge a role for electronic publishing in the field. This article provides a framework to help editors and scholarly societies develop electronic publishing policies.

This policy homogenizes all forms of Internet posting ("putting on the Internet"), and declares them categorically equivalent to publishing.  Even the current APA policy  reflects this view.  Posting, however, could include a range of activities, such as posting a document:

- on a password-protected Web page in order to share with co-authors
- to a private departmental e-mail discussion list for comments
- to a public e-mail discussion list for comments and feedback
- on a personal home page
- on an institutionally-maintained working article or pre-print server
- by a peer-reviewed electronic journal with no paper counterpart

Each of these different posting practices exhibits different communicative properties: different audiences, different restrictions on readership, different representations of the status of the document.

Postings could be made available for various time intervals, such as:
1. A document could be made available on a password-protected Web page for a short period, in order to get comments, and then be taken down.
2.  A document could be made available on a personal home until acceptance or publication in a paper journal.
3. The same document could be made available for a year or more, even after it is published in a paper journal.

Postings may also be in any one of a number of data formats, such as ASCII, Rich Text Format, HTML, TeX, PostScript, or PDF.  Certain formats, such as PDF, HTML, and PostScript may allow richer documents to be posted, but may make more demands on the user to retrieve, view, or print properly.

Some of these formats may be more easily readable  by the participants of certain communities and in certain locations than others.  For example, a PostScript document could probably easily be printed in a physics laboratory, in which the researchers have ready access to PostScript printers, UNIX workstations, and are used to working with PostScript.  That same document may be essentially unreadable in a small, low-budget humanities department.  Further, the author of the document may not be directly responsible for the format in which it is has been posted.  For example, a working article series or conference proceedings editor may reformat documents into a common format, such as PDF, for consistency.

All of these activities are equivalent, however, in the 1996 APA policy.  Ironically, the APA, along with other scholarly societies, recognizes a continuum of activities in the print world, only some of which would be considered "publishing." For example, a printed article could be distributed by being:

- Distributed by hand via a stack of copies in the authors' office
- Handed or mailed 30 copies out to trusted colleagues
- Included in a course packet
- Distributed as handouts at a talk or conference with no published proceedings
- Distributed by mail as a pre-print or working article to interested colleagues
- Distributed in a language other than the original language of the article
- Published in a regional journal or conference proceedings
- Published in a high-impact journal in a field

It is unlikely that a scholarly society would consider documents distributed in all of these ways to have been "published"; they recognize heterogeneity of distribution practices in the print world. Yet the 1996 APA policy declares all documents posted on the Internet to have been equivalently published.

The APA's 1996 embargo on Internet posting is not anomalous; the American Chemical Society (ACS) has a similar policy with respect to the *Journal of the American Chemical Society*:

> "As stated in the Notice to Authors of Papers submission of a manuscript to the Journal implies that the work reported therein has not received prior publication and is not under consideration for publication elsewhere in any medium, including electronic journals and computer data bases of a public nature. The editors and the advisory board have established a policy that any material that is posted in electronic conferences or on WWW pages or in newsgroups will be considered as published in that form, in the same way as if that work had been submitted or published in a print medium (American Chemical Society, 1996)."

These policies stand in contrast to common practices of some other fields, such as computer science and particle physics. For example, the Association for Computing Machinery's (ACM) interim copyright policy, for example, does not homogenize all forms of posting on the Internet, nor does it declare the posting of a document at any stage of development as equivalent to publishing. In part, the ACM's Interim Copyright Policy states that:

> "ACM intends to be the author's agent in reaching the widest possible readership and protecting the author's interests against plagiarism and unauthorized copying or attribution of an author's work. The ACM grants authors liberal retained rights including unlimited reuse of the work with citation of the ACM publication and the right to post preprints and revisions on a personal server (ACM, 1995)."

ACM, thus, sees its role not as the sole provider of a work, but rather as a facilitator of wide readership access, and maintainer of the "version of record" of the author's work. Only the "definitive," published article need be maintained on the ACM Web server.

The ACM policy parallels the practices found in the particle physics community as well. When an author submits an article to a journal for publication in particle physics, the author typically posts the document *at the time of submission* on one of several publicly available working article (or "e-print") servers. This document is then available for others to read, even before it has been received or reviewed by the journal.

*MIS Quarterly*, a high impact journal in the Information Systems field, represents yet a different practice. Authors of articles that have been accepted for publication in *MIS Quarterly* may post article drafts on their own Web pages, with explicit notice that these drafts are "pre-prints," and are thus not the official, "published" versions of the article. Links to these drafts are collected on a Forthcoming Articles page on the MIS Quarterly Web site. However, once the journal issue is available, the author's draft must be taken down from the author's personal Web page (MIS Quarterly Web Site, 1998).

*When Does Internet Posting Facilitate Scholarly Communication?*

The final, and most important, tension motivating our analysis is the pivotal issue of whether electronic publishing prior to peer review can improve scholarly communication (through the selective use of electronic publication) without undermining the communication currently provided by journals. There is a common assumption that posting a document on the Internet enables rapid access to a large percentage of probable future readers within the relevant scholarly communities for that document. This assumption is implicit in the APA and ACS policies: if posting did not ensure a wide readership, then posting would not pose a threat to the copyright of APA or ACS. However, the assumption is frequently voiced more explicitly, as it is, for example, in a recent press release about electronic publishing from USACM:

> "Scholars will have to choose between being widely read and being peer-reviewed."

or in the publisher's introduction to The Internet Journal of Science, a pure electronic journal in Biochemistry:

> "We believe that scientific publishing will be revolutionized by the Internet and change from a very elite medium to one accessible by anyone on the planet. (http://www.netsci-journal.com/docs/publish.htm)"

The key concept here is that of the primary scholarly community. Put differently, getting a document on the Internet read by the right audience takes work. Academics, like most professionals, are busy people, and many do not go out of their way to comb the Internet for possibly relevant material. As one astrophysicist informant put it, "astrophysicists don't surf the Web for astrophysics (Kling and McKim, 1997)."

Posting a document in an unrestricted site on the Web potentially expands its readership to millions of people for little or no marginal cost. But a document's availability on the Web does not mean that it will be read widely by the relevant community. In order to understand this, we must also consider the degree to which and how communications within a scholarly community are structured by institutional circuits.

## Policy Questions

These three tensions lead to two policy questions that scholarly societies and journal editors must face with respect to electronic publication and scholarly communication. First, what scholarly publication status should be given to electronic documents that are posted on unreviewed Web sites, such as a personal home page or a working article server? Second, when should journal editors consider electronic documents to have been previously published, in a manner that should prohibit subsequent independent publishing of the same document? In examining these policy questions, we also provide a framework with which to analyze claims made about the relationship between various forms of paper and electronic publishing.

These issues go beyond policy questions for journal editors and scholarly societies, however. The perceived scholarly publication status of electronic documents posted on the Internet plays a vital role in scholarly behavior, both with respect to the types of documents that scholars are willing to post on Web sites, and in what forms, as well as the trust with which scholars will perceive documents posted on the Internet in various forums.

It can be helpful to characterize two potential scenarios that journal publishers and scholarly societies often fear. The first can be called *the eroding subscribership* scenario. As more scholars self-publish via the Internet, other scholars gradually find that they can get the articles they want as self-published documents, in some cases months or years prior to official journal publication. Gradually, as the proportion of articles that scholars want and can readily locate on the Internet increases, they cancel their journal subscriptions. Eventually, the journal ceases to be economically viable. This scenario is unlikely for several reasons, including (1) many scholars value the quality and filtering of peer-review, (2) many journal subscriptions come as a part of a society membership, and (3) many find the journal to be a convenient package and do not have time to search the Web routinely.

An alternative, much more plausible, scenario, is that of *the electronic aggregator*. As scholars self-publish more articles on the Web, links to these documents are collocated by a third-party. Scholars might then begin using these aggregation pages as a surrogate for journals themselves. This scenario could develop in several directions. First, the electronic aggregators could provide their own peer-review function themselves (or, more likely, farm it out to members of their community). This scenario is similar to that described in Harnad's "subversive proposal." Then, the aggregation becomes, in essence, an e-journal.

Second, the aggregators might only collect and link to drafts of articles that have been accepted for publication in an existing peer-reviewed journal. For example, *MIS Quarterly* (*MISQ*) allows authors to post drafts of upcoming articles on the Web. If those articles weren't required to be removed upon publication, an electronic aggregator could simply read the tables of contents for upcoming journal issues, find the drafts of articles that will be published in *MISQ*, and create a "shadow-journal" – a Web site with links to draft-versions of articles published in *MISQ*. The degree to which a shadow-journal might erode subscribership to the journal that it shadows could depend on several factors, including the degree to which articles are revised during the review process, and the value of other features provided by subscription (such as access to archives and other value-added materials).

The ACM's "Interim Copyright Policy" mentions this particular scenario as a potential threat:

> "Someone who creates a work whose pattern of links substantially duplicates a copyrighted work should get prior permission from the copyright holder. For example, the creator of "A Table of Contents for the Current Issue of TODS" -- consisting of citations and active links to authors' personal copies of the articles in the latest issue of TODS -- needs ACM permission because that creator is reproducing an ACM copyrighted work. If all the links in the "Table of Contents" pointed to the ACM definitive versions, ACM would normally give permission because then the new work advertises an ACM work. To avoid misunderstandings, consult with ACM before duplicating an ACM work with links (ACM, 1995)."

The electronic aggregator scenario could also evolve into what Cameron (1997) proposed as a "universal citation database."

*Publication Status of Electronic Documents*

Our key question is: what publication status should scholars give to electronic documents, both relative to each other, and relative to various forms of paper documents? The concepts of publishing and relative publication status are rarely examined in discussions of electronic publishing [5]. Posting on the Web is frequently and casually compared to publishing both in professional and academic discourse. Some Web editing software programs, such as Netscape Composer, refer to the function that allows the user to make their documents available on a Web server as "publishing" (i.e. "One-button publishing"). News stories frequently claim that anyone can be a publisher, on the Web.

Dictionary definitions of "publish" are also revealing:

> "Middle English, modification of Middle French *publier*, from Latin *publicare*, from *publicus* public
> 14th century
> 1 a : to make generally known b : to make public announcement of
> 2 a : to disseminate to the public b : to produce or release for distribution; specifically : PRINT"

There are certain sensitizing situations in which academics (and others) are forced to examine their own publishing practices. When authors of a report decide between potential publication forums, they consider which forum will best reach the appropriate audience, as well as which one will best advance their careers (Rabinow, 1996). Researchers working at the intersection of multiple disciplines with significantly different and contradictory publication practices (such as psychology and computer science) may frequently need to negotiate conflicting norms.

---

[5] The recent book by Crawford, Hurd and Walker (1996) is an exception. They extend Garvey's (1979) model of scholarly publication trajectories (talk, conference paper, journal article) to include electronic media. But they don't theorize how these different forms of scholarly communication differ.

Scholarly communities have developed conventions about the relative status of different paper documents. This is reflected in highly differentiated category systems of books, journals, reports, conference proceedings, working papers, and field-specific valuations of these documentary formats (and publishing venues). For example, while both computer-science and biology rely upon conferences extensively, computer-scientists value conferences as a final publishing forum. In contrast, biologists typically do not, viewing them as a more informal forum for sharing results. Many humanities disciplines, such as literature and history, value books as a publication forum, while the lab sciences typically devalue book and book chapter publication. In many areas of physics, talks hold a high formal status (Riordan, 1987), while most disciplines use talks primarily for informal communication.

However, as the policies of the APA and ACS reveal, the relationship between various forms of electronic publishing and various forms of paper publishing have not yet stabilized, either in society at large, or in most academic communities.

## Characterizing Scholarly Publishing

*Scholarly Publishing as a Communicative Practice within a Community*

In much academic discourse, publishing is treated from an implicitly functionalist perspective. That is, scholarly publishing is discussed with respect to the functions that it fulfills within a scholarly community – generally to communicate results, allocate status, and allocate resources. In this analysis, we treat scholarly publishing as a **communicative practice** – an activity engaged in by scholars who primarily want their reports to be widely read and credited by their target audiences[6]. However, it is also essential to view this communicative practice as being anchored in a particular community (or communities, as is often the case for scholars working in such interdisciplinary fields as urban studies and gerontology.) Making this distinction is crucial in being able to move beyond false dichotomies such as this one, which appeared in a recent press release about electronic publishing, and is characteristic of the conventional view that equates Web posting with publishing:

> "Scholars will have to choose between being widely read and being peer-reviewed."

The common dictionary definitions of "publish", "a : to make generally known b : to make public announcement of" are implicitly indexed to the author's scholarly community. To publish an article, based on this definition, an author would need for their article to be announced to a substantial fraction of her scholarly community. An academic who "publishes" a scholarly article by leaving 10 copies of the articles in every county courthouse in Indiana would not be taken seriously by her colleagues. When an author is deciding on the appropriate forum for publishing a report, part of the assessment of various outlets and formats (i.e. journals, edited books, anthologies, etc.) is based on an element of quality – but also based on effective publicity within a scholarly community. Trade and academic publishers often rely upon formal criteria with respect to prior publication – they will generally not republish material that has already been published in a previous journal or book. However, business judgments are also influential, as, for example, in purchasing paperback and reprint rights to a work[7].

---

[6] We do not discount the importance of the functionalist model; in fact, the perceived status differences between publication venues as viewed by academic search and screen committees, tenure and promotion committees, grant review panels, and departmental chairs and deans plays a major role in selection of publication venue by a scholar. We are simply providing another perspective – one that is based on scholarly publishing as a communicative act.

In order to examine the relationship between being published and being read, we also need to examine access to electronic materials. We cannot assume that access to a document via electronic means is equivalent to access to the same document in a paper form, with respect to the degree to which accessing to the document fits into the reader's work style. Many people find paper media, particularly the paper journal to be a remarkably convenient format for noticing new articles, for storing them, and for reading them[8].

*When is a Document Effectively Published?*

We conceptualize publication, as a multidimensional continuum, rather than as a discrete, binary category, and one that is anchored in a particular field[9]. When a scholarly document is effectively published within a scholarly community, it seems to satisfy three criteria: **publicity**, **access**, and **trustworthiness**. Figure 1 summarizes these three criteria.

*A Model of Scholarly Publishing*

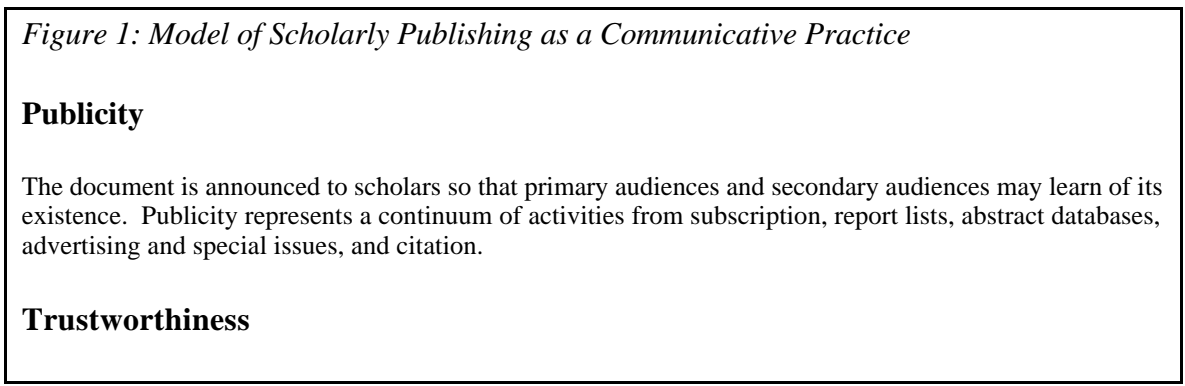

*Figure 1: Model of Scholarly Publishing as a Communicative Practice*

**Publicity**

The document is announced to scholars so that primary audiences and secondary audiences may learn of its existence. Publicity represents a continuum of activities from subscription, report lists, abstract databases, advertising and special issues, and citation.

**Trustworthiness**

---

[7] Books, for instance, are occasionally reissued by a second publisher, often as part of a series.

[8] This is not to discount the genuine limitations of the journal format, particularly as the number of papers published by a particular journal, and therefore page counts, increase quickly. One of our informants, a biochemist, reported using the electronic version of the Journal of Biological Chemistry in place of the paper version, at least partially because the ever-growing journal exceeded the capacity of his bookshelves!

[9] We recognize that there are times in which a binary decision as to whether a document has been published or not must be made. These situations include priority disputes, patent or other intellectual property claims, and libel claims.

> The document has been vetted through some social processes that assure readers that they can place a high level of trust in the content of the document based on community-specific norms. Trustworthiness is typically marked by peer review, publishing house/journal quality, and sponsorship.
>
> **Accessibility**
>
> Readers must be able to access the document independent of the author, and, in a stable manner, over time. Accessibility is typically assured by institutional stewardship as practiced by libraries, publishing houses, clearinghouses, and is supported by stable identifiers, such as ISBN and ISSN.

Each of these three criteria suggest a family of heuristics that scholars can use in assessing how effectively an article or book has been published within a scholarly community. Each of these three dimensions are polyvalent. For example, a work may be publicized by the author (by noting its availability in visible forums), by a producer (a journal may advertise upcoming issues), by an aggregator (by including a particular publication in an index or abstracting service), or by the consumer (through citation in a visible forum). Accessibility may be different, depending on how long after publication the reader is attempting to access the document, from the reader's location (home, the field, the office, the library), and the reader's institution (major research university, urban university or college, remote rural college).

*Trustworthiness*: Peer review is a particular form of vetting that is distinctive of the academic communities. However scholars use other signs to assess the value of a document as well, often in combination – such as the *reputation* of a journal or publishing house as indicators of reliability. Peer review practices vary across the disciplines: some social science journals rely upon double-blind reviewing; many journals seek two to three reviews, while others (the *Astrophysical Journal*, for example) assign one reviewer to each article. Book publishers vary in the level of detail in a proposal that they require for review (from a short proposal through sample chapters to a full manuscript), and in the number of reviews. At the lower end of a scale of Trustworthiness lie practices such as self-publishing, publishing in non-reviewed (or weakly reviewed) outlets (such as the working paper series of an academic department), or publishing in edited (but not refereed) journals. Even in non-reviewed or weakly reviewed venues, the reputation of the author (as perceived by the reader) may be a major factor in determining trustworthiness.

This analyses of trustworthiness refer to institutionalized practices that are "beyond the person." Each scholar knows others whose works s/he trusts and would be eager to read in a prepublication form. But these judgements rests on a mix of highly personal knowledge, tastes, and interests.

We see no in-principle difference between e-media and (paper) p-media regarding the trustworthiness of articles. In practice, there are (pure) electronic journals that peer-review articles for possible publication, and paper media – such as newsletters and bulletins that are lightly edited. While few high-status scholars may currently publish in e-journals, drawing a conclusion from this would be specious, since there are vastly fewer e-journals than print journals, and few e-journals have been around more than a couple of years.

*Publicity***:** A book or article is more effectively published to the extent that members of its primary and secondary audiences are made aware of its availability. Today, articles are publicized when readers see a copy of the publication that contains the article – primary publicity and access to the article are coextensive. This contrasts with publicity about books where publishers rely upon book catalogs, print ads, and published reviews as major publicity media, but the reader does not have access to a copy of the book at the same moment as receiving an announcement of the book's existence. Books can also be announced through promotions via book stores and libraries ("new book" shelves), which do more tightly couple publicity and access.

Articles and books (to a much lesser extent) are also announced to their readership by appearing in abstract indices and agglomerations (such as PMLA, Chem Abstracts, MedLine, or a Dialog database). This form of announcement is "demand driven" since potential readers are usually searching for potentially relevant publications by topical indicators.

In principle, e-publication (such as posting on a Web site or in a forum on the Web) would seem overwhelmingly more likely to effectively advertise a book or article when compared with publishing in a paper journal, or surpass the relatively limited efforts of many (paper) book publishers to advertise their wares. In practice, the differences are more subtle, since relatively few established scholars regularly read (pure) e-journals or seek them out, and many book publishers are attempting to exploit the Internet as a publicity medium. Further, many Web sites are "weak attractors" of reader interest. A major paper journal, which a well-established readership and reputation (e.g. *Science*, *Nature*) may be able to publicize the results of a study within a particular readership community far more effectively than a typical Web site.

*Accessibility*: Central to the notion of being effectively published is a perception that an author's work can be readily located and obtained by interested scholars. The improvements in the communication infrastructure of Interlibrary Loan in the last decade has increased the effective accessibility of books and articles. Even so, the pejorative notion of "the obscure journal" refers to one that few scholars could locate or find.

At the very least, to be accessible by (most) scholars, a scholarly publication has to have stable identifiers (such as author, publisher, date and an ISBN number, or be published in a journal with an ISSN, and volume and date identifiers).

In addition, scholars want materials to be "entered into the record" for a long time. While terms such as "permanent record" and "in perpetuity" indicate the moral time frame for scholarly access, more modest aspirations would probably satisfy most scholars – ranging from perhaps decades for many physical scientists to between two and three centuries for many scholars in the humanities.

Accessibility is the bane of many documents published e-media. While the physical reproductive medium may last decades, many digital materials can become unreadable as recording technologies change and current technological formats become obsolete. It is hard to find an 8" diskette drive, or a program that will faithfully translate documents written in Wang 30 word processors, Apple II word processors and so on.

Networked systems, such as the Internet, are not archival media without significant human stewardship. Today, it is common for faculty and departmental home pages to be moved from one computer address to another as local computers are replaced or reconfigured. Scholars may post documents on their home pages, but their institutions have no responsibility for the stewardship of their digital corpuses when they leave or die. Today, it is common for an active searcher for documents on the Web to find many indications that whole sites (including journals) have moved to another URL; and messages about broken links and non-existent or inaccessible pages are commonplace (Koehler, 1999)[10].

Long term, stable accessibility requires active stewardship and is more reliable to extent that stewardship is institutionalized (Task Force on Archiving of Digital Information, 1996). The existing paper-media-based scholarly communications system fulfills both of these criteria. Research libraries provide long-term access to their holdings. Their stewardship is strongly institutionalized and it does not depend upon the goodwill of a volunteer. Even Paul Ginsparg recognizes the importance of trans-organizational, institutionalized stewardship in maintaining access:

> "The physical format, with a worldwide system of institutional libraries serving as a multiply redundant distributed archive, has proven robust on the time-scale of centuries to anything short of global cataclysm (in which case we would probably have more pressing concerns) (Ginsparg, 1996)."

The institutionalization of document stewardship is facilitated by shared standards, classification systems, cataloging procedures, and professional practices. (There are, of course, well known limitations in the current library system because of gaps in each libraries holdings, delays of interlibrary loan, and important materials that librarians cannot or will not circulate.)

An organization could create an institutional framework for stewardship of an electronic corpus, for example, by creating a group with a charter and an endowment to preserve a specific e-media corpus "in perpetuity." Such a group would be staffed and paid in a way that is loosely similar to an endowed library – with archivists, preservation specialists, etc. While the technical skills

---

[10] Accompanying electronic publishing enthusiast Andrew Odlyzko's essay "Tragic Loss or Good Riddance? The Impending Demise of Traditional Scholarly Journals" (Odlyzko, 1996) are a set of instructions for obtaining preprints of the essay via e-mail from an AT&T document server. Unfortunately, the document server has apparently been reorganized, to the point where following the instructions result in the following error message: "Mail to `research.att.com!netlib' alias `local!netlib' from 'indiana.edu!mckimg' failed. The mailer `/v/lib/upas/route.reject 'netlib'' returned error status 1. The error message was: User 'netlib' is unknown at research.att.com." This phenomenon is unfortunately experienced all too frequently by users of electronic services

would be considerably different than those required to carry out functionally similar activities with paper media, the stewardship responsibility would be similar. Further, the more that such stewardship is institutionalized and embedded in professional practice, the more successfully stable access will be maintained. However, this conception is at variance with the common ideology of most e-media enthusiasts that e-media are inexpensive and uncomplicated for scholars to use, anytime, anywhere (Covi & Kling, 1998).

## When Does Posting a Document on the Web Constitute Effective Publishing?

We will employ our three criteria for publication as a communicative practice – publicity, access, and trust – in order to analyze a series of common distribution practices, both in print and on the Web. After analyzing three cases, self-publishing on the Web, self-publishing in a pure electronic journal, and publishing a dissertation on an institutionally maintained dissertation server, we provide a table enumerating a series of additional cases – different forms of posting and article distribution – whose "strength of publication" can be assessed with this framework. This analysis will show that the "Posting is publishing" model equates practices that constitute very effective publishing in the paper world with practices that do not *at all* constitute effective publishing in the electronic world.

*Self-Publishing on the Web*

This form of self-publication refers to posting a document on a personal home page, with no active stewardship, and no additional announcement.

> *Publicity:* Self-publishing on the Web, particularly on a personal Web site, generates little publicity. Auxiliary publicity-generating activities, such as posting an announcement of a document's availability on an e-mail distribution list, or having the availability of a document noted in a bulletin or newsletter, the self-posted document on a personal Web site is unlikely to be noticed by one's peers. Of course, this can change over time. A document may be indexed by search engines and thus be returned in query results, although many universities block search engines from indexing parts of their sites. The document may be linked from other pages or resource guides.
>
> *Trustworthiness:* The trustworthiness of a self-posted Web document depends almost entirely upon the author's reputation within a particular scholarly community. For example, a non-peer-reviewed posting on a Web site by a high-status and well-respected scholar may well be trusted more than a peer-reviewed journal article by someone not well-known in the community.
>
> *Accessibility:* Accessibility also changes over time. For example, accessibility of a document on the Web in the short run is very high; anyone with an adequate Web browser and Internet connection can access the document independently of the author. However, over time, the document may become inaccessible, without active stewardship.

> For example, if a Web site address is changed, as often occurs, the document will be less accessible unless appropriate redirections are created, search engine entries are updated, and so on. When the author of the document leaves the institution, either through moving (as happens frequently with faculty and always with graduate students) or retirement or death (as happens eventually), then the address of the document will change as well. Therefore, without active and careful curating (which is cognizant not only of the document and its status, but the means through which it is being found and accessed), the document posted on a personal Web page will become less accessible over time.

These considerations lead us to view self-publishing on a personal web site to be a very weak form of publishing.

However, in conjunction with publication in another forum (e.g. a journal or a conference), posting a document on a personal or organizational Web site may perform a useful service both to potential readers of the document and even, in certain cases, to the forum itself.   First, posting a document on a personal Web site in advance of publication in another forum may serve as a convenience to those who are already likely readers of a particular document – colleagues and other members of the author's invisible college.  This is akin to sending out pre-prints of an accepted article.  In the short run, accessibility may actually be substantially increased, at least to members of the author's invisible college.  Second, a document posted on a Web page in advance of publication in a journal in which it has been accepted may actually serve to lead potential readers to, and therefore advertise, the journal.  This is particularly true for a lower circulation journal, with a relatively stable and well-defined audience; an article posted on the Web by the author may actually serve to attract additional audiences to the journal[11].

*Publishing in an E-Journal*

In contrast with  a document being self-published on a personal Web page, consider the example of a document being published in a pure e-journal.

> *Publicity*: It may benefit from greater publicity, as most electronic journals attempt to maintain a systematic subscriber list (even if the journal itself does not cost anything to access), and some pure electronic journals may even eventually be indexed by various indexing and abstracting services (as is *Public Access Computing Systems Review*, for example).
>
> *Trustworthiness:* The document's trustworthiness depends upon the journal's quality reputation  readers. Today, many e-journals are struggling for legitimacy and some of them will  improve over time.  They face both the struggles of all new journals to develop a reputation for quality, as well as having to overcome some of the current stigma of also being a pure e-journal.

---

[11] The potential benefit the journal may receive might or might not outweigh any loss the journal might suffer as a result of readers benefiting from the vetting and filtering functions performed by the journal but not subscribing.  This would be a matter for empirical investigation.

*Accessibility:* The accessibility of documents over time depends upon stewardship of the journal, including their long-term funding. Stewardship is greatly enhanced by the involvement of an institution and institutional apparatus, rather than merely an individual, in addition; for example, access to a pure electronic journal run by a scholarly society may be more stable over time than one run by an individual or a small unaffiliated group of individuals. In addition, an e-journal may be more likely to actively collected (catalogued, even archived) by a library than an individual, self-published Web document. The involvement of the library and its procedures and practices greatly enhances the stability over time. The publisher's introduction to *The Internet Journal of Science: Biological Chemistry* fails to recognize the importance of institutions in maintaining stable access over time:

> "Of course the costs of providing such information to you the reader are also reduced because there is no need for a local library to store the information (http://www.netsci-journal.com/docs/publish.htm)."

Stable access does not require that every library maintain its own electronic copy; in fact, too many Web sites of public availability may actually be confusing to researchers, and may also exacerbate the version problem. However, some redundancy and institutionalized stewardship facilitate stable access over time.

*Publishing in the Virginia Tech Electronic Thesis Server*

In 1997 the Virginia Polytechnic Institute and State University proposed that all of their graduate students post their masters and doctoral theses in an electronic archive, the Electronic Thesis Server (ETS). The ETS is available on the Web at no cost to readers (*New York Times*, 1997). Although other universities are participating and contributing to the archive, Virginia Tech is the only university to propose that it be required for its graduates. This case is controversial and interesting because some fear that such posting of dissertations on the Internet constitutes a form of prior publication that may make it more difficult for graduate students (and faculty advisors) to publish their dissertation research later in journals. One recent chemistry graduate commented that "The problem is that most of the chemical journals will not take something that's already put out on the Net (*New York Times*, 1997)." Students could thus be caught in a bind in which the dissertation is not considered to be effectively published by universities making hiring and promotion and tenure decisions, while it is considered too effectively published for publishers to allow republication (even in a revised form, such as an article).

We can use our framework to examine whether publishing dissertation as part of the ETS has been so effective that it could preclude further publication. The effectiveness of publication through ETS should probably be compared to the traditional channel which dissertations are published: UMI.

*Publicity:* The publicity of dissertations published in ETS is fairly weak, although no worse than with dissertations published through UMI. Dissertations are rarely publicized systematically in research communities. This is one reason why many publishers have been willing to republish in a revised form research originally disseminated in a dissertation.

*Trustworthiness:* The criterion of trustworthiness is certainly met, in the sense that the trustworthiness is determined by the evaluation procedures of the university, and that legitimacy is marked by the approval of the dissertation by the university graduate school and the author's committee. This marking is included in the ETS. Thus there is no reason to believe that the trustworthiness of the dissertation is any less (or any more indeterminate) than any other dissertation, available in paper form from the UMI dissertation clearinghouse.

*Accessibility:* Improved access makes the dissemination of a dissertation through ETS potentially a stronger form of publication. First, access to dissertations posted on ETS is highly institutionalized, and therefore author-independent. Second, Virginia Tech has developed and implemented institutional structures that can maintain stable access over time. The creation of the archive was funded by a United States Department of Education Grant. Upon submission of a dissertation into the archive, students pay a fee (in lieu of the traditional dissertation binding fee) that is used to maintain the archive in perpetuity. Finally, these electronic dissertations are also automatically submitted to UMI, who then makes paper archive copies which are then filed and distributed like any other UMI archives. Because of this stewardship, which includes a funding model and funding sources (binding fees) and archiving (through submission to UMI), access may well be quite good, and remain stable over time.

In summary, the electronic forum, ETS, actually constitutes a more effective form of publication than the paper equivalent. Ironically, precisely because of fears that access was so good that having a dissertation published via ETS would preclude further publication, the managers of ETS now allow authors to specify that their dissertations are available only from within the Virginia Tech campus network, a form of publishing closer to putting a copy of the dissertation on reserve at campus libraries, and thus weaker than traditional publishing of a dissertation through UMI. Thus, in the case of ETS, publication via the Internet can be seen to constitute a form of publication at least as effective as the equivalent paper publication of a dissertation through UMI, and potentially more effective in several ways.

**Posting a Document on the Web: Illustrative Examples**

The following tables characterizes a series of Web posting practices, with the degree to and direction in which they move posting along the three publishing criteria. The first table (Table 1) represents "simple posting practice", or self-publication – posting a document on a personal home page, with no active stewardship, and no additional announcement. The putative effects of this practice on publicity, access, and trustworthiness are illustrated at three points in time.

*Table 1: Simple Posting Practice (Self-Publishing), 3 Time Frames*

| Practice | Publicity | Access | Trust |
|---|---|---|---|
| **Simple posting, after 1 month** | No publicity -- report has not been indexed by search engines, probably hasn't been substantially linked-to. | Good access, if reader knows where to look. | Dependent upon reputation of author. |
| **Simple posting, after 1 year** | Increased publicity -- page has probably been indexed by several Web search engines. | Increased, as search engines index documents. | Dependent upon reputation of author. |
| **Simple posting, after 5 years** | Increased – page has been indexed by Web search engines. | Decreased – without active maintenance, it is likely that links have decayed, documents have been moved, etc. | Dependent upon reputation of author. |

Tables 2 considers a series of modifications of the simple posting practice illustrated in Table 1. The effects of each of these practices on publicity, access, and trust should be considered in comparison to the simple posting practice described above.

*Table 2: Posting Practices on a Personally Controlled Web Space*

| Practice | Publicity | Access | Trust |
|---|---|---|---|
| **Personal Web Space:** Posting in personal Web space | May increase, only if person's Web page is frequently checked by others in the community, and seen as a useful resource | May be decreased, particularly over time. People move (students always, faculty often), servers are changed or reorganized. | May be decreased, unless poster is already well-known, well-established, and well-respected. |
| **Self-announcement on Listserv**: Posts notice of report's availability on one or more listservs | Increases, depending on the size and character of the readership of the listservs (e.g. the degree to which the listserv's readership is representative of the research community) | | |
| **Recommendation on Listserv:** Notice of report's availability posted by someone other than the author, in the form of a recommendation. | Increases, depending on the size and character of the readership of the listservs (e.g. the degree to which the listserv's readership is representative of the research community) | | Increases or decreases, depending on the reputation of the recommender. |
| **Bulletin Notice:** Notice of posting in a paper bulletin or newsletter. | Increases, particularly if newsletter is widely disseminated within a community (a society bulletin, for example) | | May increase if bulletin is well-respected or represents a scholarly society |

*Table 2, Continued*

| Practice | Publicity | Access | Trust |
| --- | --- | --- | --- |
| **Password Protection:** Posting on a password-protected Web site | | Substantially decreases – more akin to an author personally disseminating papers to trusted colleagues or co-authors | |
| **Site Restriction:** Posting on a site-restricted Web site (e.g. Virginia Tech thesis server) | | Substantially decreases – similar to a printed copy being available only in the reading room of one library | |
| **Resource Guide Link:** Link from an electronic disciplinary directory or resource guide (a la Pedro's Tools) | May be substantially increased, depending on the centrality/frequency of use of the disciplinary directory | May be substantially increased, particularly if the disciplinary directory is actively curated – if the URL of the original document changes, the curator may attempt to control for that change by updating the link. | May be increased, depending on the reputation of, and the selection criteria used by, the maintainer(s) of the disciplinary directory |
| **Robot Blocking:** Posting on a robot-blocked Web site | Decreases – if Web search engines are unable to index report, it will likely not be included in the results of a search query | Decreases – readers will need to have the exact URL or more direct pointer (and remember this over time) if document is not available via Web search engines | |
| **Undescriptive Header:** Posting without adequately descriptive title or page header | Decreases – Web search engines may be less likely to include the document in the results of a query, and, if they do, may return it with a lower relevance ranking | Decreases – may be harder to find document if one doesn't have the exact URL | |

*Table 3: Posting Practices on an Institutionally Controlled Web Space*

| Practice | Publicity | Access | Trust |
|---|---|---|---|
| **Institutional Web Space:** Posting in institutional/organizational Web space | | May be increased, particularly if institutional space is well-curated. | May be increased, if institution is well-established and respected (example: Center for Coordination Science at MIT) |
| **Working Article Server:** Submission to a working article server, such as http://www.arxiv.org/ | May increase, particularly if working article server is visited frequently | Increases, particularly as working article server is curated and funded. | |
| **Electronic Archive:** Inclusion in an electronic archive or collection | | Increases with stewardship practices, such as cataloging, format preservation, backups, etc. | Little effect within a research community per se – may have some effect on more peripheral communities (students, general public) |
| **Pure Electronic Journal:** Publication in a journal distributed primarily in electronic form | Depends on the subscription and readership of the journal. Most pure electronic journals at the moment have very small readership. | As with the electronic archive, increases with stewardship practices. An electronic journal started without significant institutional support may fail, and without alternate archiving arrangements, access may decay over time. | |
| **Hybrid (P-E) Electronic Journal:** Publication in an electronic version/edition of a journal distributed primarily in paper form. | Depends on the subscribership – however, many hybrid journals, particularly high-impact journals may essentially cover the entire research community. In addition, the print and electronic announcement may reinforce announcement. | Greatly increased – the ease of access may be facilitated the availability of the journal in electronic form, while the print journal, along with its integration into library and other institutional practices may ensure stability over time. | |

Finally, in Table 4, we present, in a similar format, a series of print publishing practices. The table illustrates how different publishing practices, all commonly engaged in by scholars, are heterogeneous, and have greatly differing communicative properties. Practices towards the beginning of the table (such as distributing a few paper copies to trusted colleagues) have substantially different communicative properties than those towards the end (such as publication in a high-impact journal in the field), and few scholars would conflate the two into a homogeneous "published" category. However, many scholars and some scientific societies are indeed willing to conflate equally dissimilar practices in the electronic domain.

*Table 4: Print Publication Practices*

| Practice | Publicity | Access | Trust |
| --- | --- | --- | --- |
| **Co-Author Distribution**: Giving copies of a report to co-authors | No effective publicity. | No effective access outside of authorial control. | NA |
| **Colleague Distribution:** Handing or mailing a few (<20) copies of a report to trusted colleagues | Decreased | Minimal | Likely high within the author's own circle of trusted colleagues; nonexistent outside. |
| **Presentation/Conference without Proceedings:** Giving a talk or presentation without handouts or a published proceedings. | Could be substantial, depending on the attendance at the conference and degree to which the attendees are representative of a particular scholarly community. | Access is essentially nonexistent. | Depends on the prestige of the conference (major society conference, regional conference, etc.) and of the reputation of the presenter. |
| **Presentation Handouts:** Giving a presentation or talk and providing handouts to the attendees | Could be substantial, depending on the attendance at the conference and degree to which the attendees are representative of a particular scholarly community. | Only slightly more than presentation without proceedings. | Depends on the reputation of the presenter. |
| **Working Paper List/Clearinghouse:** Sending copies of a report to an open clearinghouse or mailing list. | Substantially increased, depending on the size of the mailing list, and coverage of the research community | Depends on the archival practices of the curator/manager of the clearinghouse -- generally minimal. | |
| **Working Paper Series:** Distribution and release as part of an institutional or organizational working paper series. | | | |

*Table 4, Continued*

| Practice | Publicity | Access | Trust |
|---|---|---|---|
| **Conference Proceedings:** Inclusion of a report in a printed conference proceedings | Increased or decreased depending on profile of conference. | Increased for a major conference, such as an ACM conference -- but decreased for many smaller conferences. Most libraries do not collect conference proceedings extensively. | Highly dependent upon the profile of the conference, and its sponsorship, as well as upon the author. |
| **Minor or Regional Journal:** Publication in a minor or regional scholarly journal. | Decreased, depending on subscribership to journals, library subscriptions, availability at research libraries. | Often very low -- third-tier journals are frequently not available at most libraries. May be restricted to particular specialty collections. | High to more peripheral participants, generally lower to more centrally positioned researchers. |
| **High-Impact Journal:** Publication in a high-impact journal. | High | High | High, but still depends on reputation of author. |

Figure 1 diagrams this fundamental asymmetry. In the "posting is publishing" model, practices that fall throughout the publishing effectiveness continuum in the electronic domain are equated only with the most effective publishing practices in the print world.

## Has It Already Been Published?

Returning to one of the motivating tensions for this analysis, we can now consider the question of when a document's being posted on the Internet should preclude its subsequent publication in a paper journal.

As a matter of policy, most scholarly journals will only accept material that is "original" or "new." However, novelty and originality should be based upon a journal's readership community. Our model of scholarly publishing as communicative practice can be used in determining whether or not a report has already effectively been published within the relevant readership community.

Consider first the case of a report self-posted on the Web, that is later submitted to a peer-reviewed journal. In general, it should not be considered to have been effectively published; access, trustworthiness, and publicity will all be relatively low. Of course, if the author has a strong reputation in the relevant readership community, many specialist readers may treat the article as more trustworthy. If the author attempts to publicize the article, then the effectiveness of publication may increase (consider, as a strong example, a scientist promoting a self-published article through a letter to the editor in *Science*). Finally, if the author submits the article to an institutionally maintained archive (such as the Los Alamos National Labs physics working article server), then stable access over time is improved. Journal editors and editorial boards should have the flexibility to take all of these criteria into account when determining whether or not a document that has been posted on the Web has already been published, from the perspective of subsequent journal publication, within the relevant readership community.

We strenuously avoid making an essentialist distinction between paper and electronic media in this argument, however. One could easily imagine another case in which an editor of an e-journal must make a decision on whether to publish a report that has already published in a paper forum. For example, an author of a report published in a conference proceedings, say, the *Proceedings of the American Society for Information Science*, might later submit the same article to *Public Access Computer Systems Review (PACS-R)*, hoping for wider distribution. The editor of *PACS-R* would now have to make a decision as to whether or not the article has already been *effectively published* to the readership community of *PACS-R*.

It is unrealistic for a journal editor confronted with a submission that has already been posted in some way on the Internet to know precisely the degree of overlap between the readership community of the original posting and the readership community of the journal. But Internet posting is not the only form of prior publicity. Scholars may present versions of an article at many seminars and specialty workshops before submitting a refined version for formal publication. Editorial boards must articulate policies that help authors to understand when they can submit a manuscript for review. We recommend that authors indicate the nature of an article's prior exposure, and that journal editorial board have discretion in determining whether or not a submitted manuscript has already been effectively published within the relevant scholarly communities.

This argument is independent of debates about authors signing away copyright when they publish in a journal (e.g. Bachrach, et al., 1998; Bloom, 1998). Scholarly societies have a strong interest in promoting effective communication within the field; in fact, this is frequently their primary function. Whether or not they require authors to sign away copyright to the article, scholarly societies have the option of viewing author posting on the Internet either as prior publication, and therefore categorically ineligible for journal publication, or as complementary to journal publishing. We view this latter stance as more conducive to scholarly communication. Since few of the activities in the pre-publishing continuum promote strong publicity, trustworthiness, and access, they pose little threat to traditional journals. Instead, journals can maintain the strength of their brand image through quality control and labeling, while allowing authors to experiment with various posting activities, in attempts to increase readership, or gain feedback prior to journal publication.

## Conclusions

Our analysis leaves us with several sensitizing concepts and normative recommendations. These should be taken into account, both by scholars considering various electronic media for their formal communication, and by scholarly societies assessing the role that the Internet can play in future services.

The first sensitizing concept is that Web posting and effective scholarly communication are loosely coupled rather than strongly related. One popular view contrasts peer-review with wide readership. When one considers the institutionalized and structured nature of scholarly communities, peer-review may enable publication in journals that are widely read within the relevant scholarly communities. Society policies that categorically equate posting a document on the Internet with publishing the document assume a strong functional equivalence, and homogenize practices that are inherently heterogeneous and contested.

The second sensitizing concept is that scholarly communication can be conceptualized as a communicative practice anchored in three dimensions: **publicity, access**, and **trustworthiness**. Publicity does not automatically and inexorably proceed from a document's availability on a global network like the Internet; rather it represents a series of active practices that are anchored in particular scholarly communities. Although access is often viewed as the way in which electronic documents are most advantaged over paper, access is also highly problematic in the electronic environment. Maintenance of stable access over time is dependent upon active, institutionalized stewardship that is embedded in institutionalized structures and professional practice. Finally, trustworthiness in itself is a highly variegated concept. Peer-review plays a large, but not the only, part in its construction . Scholars do not treat all peer-reviewed reports as equally trustworthy; rather, they rely upon a variety processes and markers, which are dependent upon everything from the structure of the discipline itself to the social networks that the readers are embedded within. This has profound consequences both for those who see electronic publishing as a straightforward technology that can be used to widen access to scholarly material, as well as for those who fear that the use of the Internet in scholarly communication will destroy much-needed journal revenues streams.

We also made several recommendations to journal editors, publishers, and societies developing Internet posting policies for authors in the field. First, we recommend that our three criteria of effective publication – publicity, access, and trustworthiness – be used to analyze claims about different forms of paper and electronic media in scholarly communication. Secondly, we recommend that journal editors consider the degree to which the manuscript has been effectively published within the relevant readership communities, before categorically rejecting manuscripts that have been made available *in some form* on the Internet. Thirdly, we recommend that journal editors continue to experiment with and seek out Internet policies that promote more effective scholarly communication by allowing authors to use the Internet to provide rapid short-term access to articles to diverse audiences, and while not jeopardizing the valuable functions provided by journals. In this regard, we support journal editors who either permit authors to post articles upon acceptance, or that actually post the articles themselves, and then remove the "pre-print" posting upon journal publication.

As the use of the Internet becomes more and more embedded in scholarly communication in many forms, scholars will face more complex choices in managing communications through electronic and paper media. We hope that the scholarly societies and commercial publishers develop guidelines that maximize effective communication through multiple media and through a deep understanding of these multiple media.